\documentclass{iau} 
\usepackage{graphicx}

\title[SXDF-ALMA deep survey] 
{SXDF-UDS-CANDELS-ALMA 1.5 arcmin$^2$ deep survey}

\author[Kohno K. et al.]   
{K.~Kohno$^1$, 
Y.~Yamaguchi$^1$, 
Y.~Tamura$^1$,
K.~Tadaki$^2$,  
B.~Hatsukade$^3$,  
S.~Ikarashi$^4$, 
K.I.~Caputi$^4$,
W.~Rujopakarn$^{5,6}$,
R.J.~Ivison$^{7,8}$, 
J.S.~Dunlop$^8$, 
K.~Motohara$^1$,
H.~Umehata$^{1,7}$
K.~Yabe$^5$, 
W.H.~Wang$^9$, 
T.~Kodama$^3$,
Y.~Koyama$^3$,  
M.~Hayashi$^3$, 
Y.~Matsuda$^3$, 
D.~Hughes$^{10}$, 
I.~Aretxaga$^{10}$, 
G.W.~Wilson$^{11}$, 
M.S.~Yun$^{11}$, 
K.~Ohta$^{12}$, 
M.~Akiyama$^{13}$, 
R.~Kawabe$^{3}$, 
D.~Iono$^3$, 
K.~Nakanishi$^3$, 
M.~Lee$^{3}$
\and 
R.~Makiya$^1$
}
\affiliation{
$^1$Institute of Astronomy, The University of Tokyo, Mitaka, Tokyo 181-0015, Japan \\ email: {\tt kkohno@ioa.s.u-tokyo.ac.jp} \\
$^2$Max-Planck-Institut f\"ur extraterrestrische Physik (MPE), Garching, Germany \\
$^3$National Astronomical Observatory of Japan, Mitaka, Tokyo, Japan \\
$^4$Kapteyn Astronomical Institute, University of Groningen, The Netherlands \\
$^5$Kavli Institute for the Physics and Mathematics of the Universe (IPMU), The University of Tokyo, Chiba, Japan\\
$^6$Department of Physics, Faculty of Science, Chulalongkorn University, Bangkok, Thailand\\
$^7$European Southern Observatory, Garching, Germany \\
$^8$Institute for Astronomy, University of Edinburgh, Royal Observatory, UK \\
$^9$Academia Sinica Institute of Astronomy and Astrophysics (ASIAA), Taipei, Taiwan \\
$^{10}$Instituto Nacional de Astrof\'isica, \'Optica y Electr\'onica (INAOE), Puebla, Mexico \\
$^{11}$Department of Astronomy, University of Massachusetts, Amherst, USA \\
$^{12}$Department of Astronomy, Kyoto University, Kyoto, Japan \\
$^{13}$Astronomical Institute, Tohoku University, Sendai, Japan \\
}

\pubyear{2015}
\volume{319}  
\jname{Galaxies at High Redshift and Their Evolution over Cosmic Time}
\editors{S. Kaviraj \& H. Ferguson, eds.}
\begin{document}

\maketitle

\begin{abstract} 
We have conducted 1.1 mm ALMA observations of a contiguous $105'' \times 50''$ or 1.5 arcmin$^2$ window in the SXDF-UDS-CANDELS. We achieved a 5$\sigma$ sensitivity of 0.28 mJy, providing a flat sensus of dusty star-forming galaxies with $L_{\rm IR} \sim6\times10^{11}$ $L_\odot$ (for $T_{\rm dust}$ =40K) up to $z\sim10$ thanks to the negative K-correction at this wavelength. We detected 5 brightest sources (S/N$>$6) and 18 low-significance sources (5$>$S/N$>$4; these may contain spurious detections, though). One of the 5 brightest ALMA sources ($S_{\rm 1.1mm} = 0.84 \pm 0.09$ mJy) is extremely faint in the WFC3 and VLT/HAWK-I images, demonstrating that a contiguous ALMA imaging survey is able to uncover a faint dust-obscured population that is invisible in deep optical/near-infrared surveys. We found a possible [CII]-line emitter at $z=5.955$ or a low-$z$ CO emitting galaxy within the field, which may allow us to constrain the [CII] and/or the CO luminosity functions across the history of the universe.  
\end{abstract}

\firstsection 
\section{Introduction}

Deep mm/submm surveys that employ single-dish telescopes with detector arrays have revolutionized observational cosmology by uncovering a new population of submillimeter galaxies (SMGs); these are dusty, extreme star-forming populations in the early universe (e.g., \cite[Casey et al.~2014]{Casey2014}). Recent extensive follow up studies on SMGs using ALMA have provided new insights into the nature of these extreme sources, such as their redshift distributions, multiplicity (revising their number counts), and source sizes (e.g., \cite[Karim et al.(2013)]{Karim2013} for APEX/LABOCA sources in ECDF-S, \cite[Simpson et al.(2015)]{Simpson2015} for JCMT/SCUBA2 sources in UDS, and \cite[Ikarashi et al.(2015)]{Ikarashi2015} for ASTE/AzTEC sources in SXDF). 
However, despite their enormous IR luminosities ($L_{\rm IR} \sim 10^{13} L_\odot$), the contribution of SMGs to the extragalactic background light (EBL), which represents the integrated unresolved emission from extragalactic sources and contains vital information on the history of galaxy formation, is rather small (e.g., $\sim$10 -- 20 \% at 1.1 mm; \cite[Scott et al.~2012]{Scott2012}). In fact, recent ALMA observations suggest that {\it faint SMGs} or {\it sub-mJy sources}, i.e., mm-selected sources with $S_{\rm 1mm} <$ 1 mJy, may account for $\sim 50 - 100\%$ of the EBL at 1.1 to 1.3 mm bands (e.g., \cite[Hatsukade et al.~2013]{Hatsukade2013}; \cite[Ono et al.~2014]{Ono2014}; \cite[Carniani et al.~2015]{Carniani2015}; \cite[Fujimoto et al.~2015]{Fujimoto2015}; \cite[Oteo et al.~2015]{Oteo2015}). However, little is known about the physical properties, such as their redshift distribution and stellar masses, of these newly discovered sources. This is partly because sub-mJy galaxies are often faint in NIR/radio at the currently achievable sensitivities. 

Here, we present an overview of the ALMA 1.1 mm (Band 6) imaging survey of sub-mJy galaxies in a contiguous $105'' \times 50''$ (1.5 arcmin$^2$) rectangular window, where deep multiwavelength data sets are available from SXDF, UDS, CANDELS, HUGS, SpUDS, SEDS, and so on (see \cite[Galametz et al.~2013]{Galametz2013} for a summary of the ancillary data).

\section{ALMA observations: strategy and data analysis}

The observed region was selected because of the richness of H$\alpha$ emitting galaxies at $z=2.53\pm0.02$ uncovered by extensive narrow-band filter imaging surveys using MOIRCS camera on the Subaru telescope (\cite[Tadaki et al.~2013]{Tadaki2013}) and covered by a 19-point mosaic in order to minimize the number of pointing, at a cost to the sensitivity uniformity within the mosaic map. This strategy minimizes the overhead fraction within the total observation time, since the overhead for the mosaic observations was expected to be rather large when the cycle 1 call for proposals was issued. 

\begin{figure}[hbt]
\begin{center}
 \includegraphics[width=1.0\textwidth]{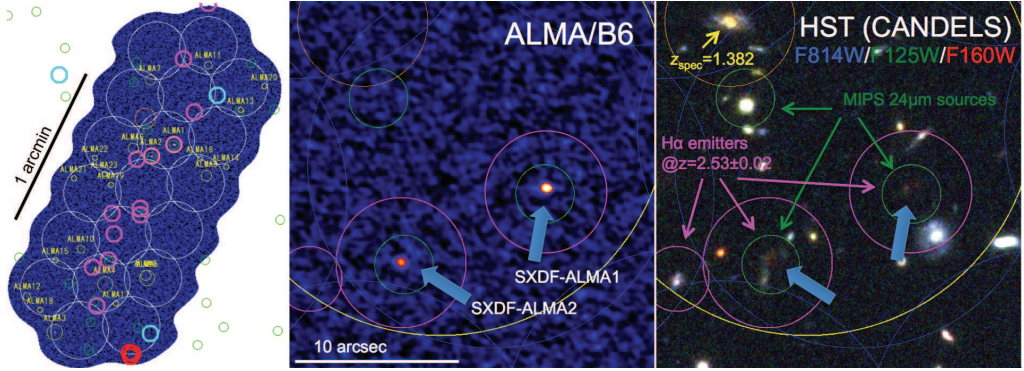} 
 \caption{(left) ALMA 1.1 mm continuum image obtained using a 19-point mosaic (indicated by white circles). Detected 23 ALMA sources are indicated by small yellow circles (larger circles denote a higher S/N) with their source ID, and the distribution of H$\alpha$ emitters at $z=2.53$ is indicated by small magenta circles. Small green circles denote the MIPS 24$\mu$m sources. (middle) A magnified view of the ALMA image showing two bright sources (corresponding to SXDF-ALMA1 and 2 in Table \ref{tab1}). (right) A 3-color composite NIR image of the same region. There are 3 H$\alpha$ emitters, 3 MIPS sources (two of these are overlapped), and 1 star-forming galaxy at $z_{\rm spec}=1.382$. We note a variety of 1.1 mm properties among these galaxies.}
   \label{fig1}
\end{center}
\end{figure}

The observations were carried out on 17 and 18 July 2014 under excellent atmospheric conditions, with a precipitable water vapor of 0.42 - 0.55 mm. Total observation time was 3.6 hours. 
During the observations, 30 - 32 antennas were available and the range of baseline lengths was between 20 m and 650 m, providing an excellent {\it uv} coverage despite the source declination (close to the equator). 
The continuum map was processed using {\tt CLEAN} with natural weighting, using {\tt CASA} package, which provides a synthesized beamsize of $0''.53 \times 0''.41$. The typical noise level is 0.055 mJy beam$^{-1}$ (1$\sigma$) near the center of each pointing. The obtained ALMA image is shown in Figure \ref{fig1}. 
We also constructed a 3D cube with a frequency width of 60 MHz in order to search for any mm-wave line emitters using a clump-find algorithm.

\section{Results and Discussion: physical properties of ALMA sources}

\underline{\it ALMA 1.1 mm continuum sources and their physical properties:} We extract significant sources above 4$\sigma$ using the {\tt AIPS} task {\tt SAD}. We detected 23 sources in total, including 5 brightest sources (S/N$>$6) and 18 less significance sources (5$>$S/N$>$4). A part of the extracted sources (with S/N larger than 4.4) are listed in Table \ref{tab1}. We estimated false detection rates by counting negative peaks in the map, indicating that sources below 5$\sigma$ may contain spurious detections, although the derived number counts are consistent with existing studies after careful estimations of completeness and false detection rate (Hatsukade et al. in prep.) utilizing a source finding code {\tt AEGEAN} (\cite[Hancock et al.~2012]{Hancock2012}). 

The multiwavelength properties of the 5 brightest ALMA sources have been studied by exploiting the rich data in this field (Yamaguchi et al. in prep.), revealing that 4 of the ALMA sources are situated on the main sequence of star-forming galaxies with a significant stellar mass ($M_{\rm star} \sim (4 - 10) \times 10^{10} M_\odot$) at their epochs ($z_{\rm photo} \sim 1.3 - 2.5$). 

\begin{table}
  \begin{center}
  \caption{A catalogue of ALMA 1.1 mm continuum sources$^1$.}
  \label{tab1}
 {\scriptsize
  \begin{tabular}{ccccccc}\hline 
ID & $\alpha$ & $\delta$ & $S_{\rm peak}$ & S/N & weight$^2$ & note$^3$ \\ 
\hline
SXDF-ALMA1 & 02 17 40.524 & -05 13 10.64 & $1.696\pm0.058$ & 29.36 & 0.952 & H$\alpha$ emitter $z=2.53\pm0.02$ \\
SXDF-ALMA2 & 02 17 41.120 & -05 13 15.19 & $0.791\pm0.065$ & 14.27 & 0.852 & H$\alpha$ emitter $z=2.53\pm0.02$ \\
SXDF-ALMA3 & 02 17 43.642 & -05 14 23.81 & $0.839\pm0.090$ &  9.29 & 0.609 & Herschel/JVLA dropout \\
SXDF-ALMA4 & 02 17 42.335 & -05 14 05.09 & $0.395\pm0.056$ &  7.05 & 0.982 & $z_{\rm photo}=1.33^{+0.10}_{-0.16}$ \\
SXDF-ALMA5 & 02 17 41.229 & -05 14 02.74 & $0.378\pm0.056$ &  6.70 & 0.976 & $z_{\rm photo}=1.52^{+0.13}_{-0.18}$ \\
SXDF-ALMA6 & 02 17 41.597 & -05 13 12.29 & $0.315\pm0.067$ &  4.73 & 0.827 & \\
SXDF-ALMA7 & 02 17 41.153 & -05 12 45.47 & $0.299\pm0.065$ &  4.58 & 0.844 & \\
SXDF-ALMA8 & 02 17 39.678 & -05 13 22.81 & $0.282\pm0.062$ &  4.54 & 0.884 & \\
SXDF-ALMA9 & 02 17 41.270 & -05 14 01.62 & $0.259\pm0.057$ &  4.52 & 0.960 & \\
SXDF-ALMA10 & 02 17 42.965 & -05 13 51.46 & $0.270\pm0.061$ &  4.44 & 0.905 & \\
\hline
  \end{tabular}
  }
 \end{center}
\vspace{1mm}
 \scriptsize{
  $^1$Only the sources with S/N larger than 4.4 are shown. \\
  $^2$primary beam correction factor. In this table, $S_{\rm peak}$ has been corrected accordingly (1/weight). \\
  $^3$source properties taken from Yamaguchi et al. (in prep.)}
\end{table}

\begin{figure}[bht]
\begin{center}
 \includegraphics[width=\textwidth]{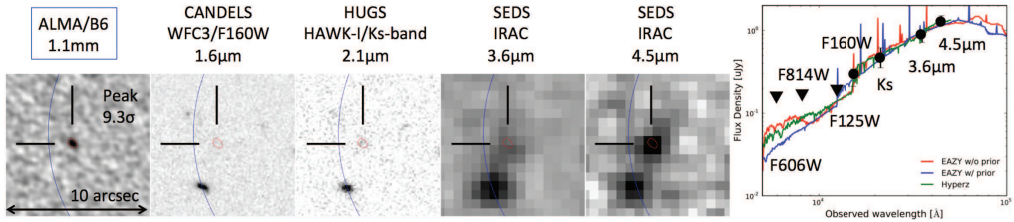} 
 \caption{Multiwavelength view of SXDF-ALMA3 together with the best-fit SED (Yamaguchi et al., in prep.). Despite the fact that SXDF-ALMA3 exhibits an elevated star-formation ($S_{\rm 1.1mm}=0.84\pm0.09$ mJy or SFR $\sim$200 $M_\odot$ yr$^{-1}$ if $T_{\rm dust}=35$ K), it is very dark even in deep WFC3/F160W (CANDELS-wide; 5$\sigma$ limiting magnitude = 27.45 $m_{\rm AB}$) and HAWK-I/K$_{\rm S}$-band (HUGS; 5$\sigma$ limiting magnitude =26.16 $m_{\rm AB}$) images. This is a ``SPIRE-drop'' source.}
   \label{fig2}
\end{center}
\end{figure}

\underline{\it New populations of galaxies unveiled by the ALMA 1.1 mm survey:} Then does ALMA only detect dust emission from ``already known galaxies'' that were selected using rest-frame UV/optical deep surveys? The answer is no; one of the 5 brightest ALMA detections, SXDF-ALMA3, is extremely faint even in the ultra-deep NIR images from CANDELS and HUGS, as shown in Figure \ref{fig2}. Further contributions to the star formation history may come from these faint submm galaxies that do not appear to be fully overlapped with UV/optical-selected galaxies (e.g., \cite[Chen et al.~2014]{Chen2014}). 

Another new type of ALMA sources is the mm-wave line-emitting galaxy; we find a promising candidate for a line-emitting galaxy at an observing frequency of $\sim$273.3 GHz with a peak flux of $3.8\pm0.70$ mJy (5.4$\sigma$) or a velocity-integrated line flux of $0.53\pm0.079$ Jy km s$^{-1}$ (6.7$\sigma$) that exhibits a galaxy-like line width (FWHM $\sim$100 km s$^{-1}$). Although it is not yet clear at this stage whether this is a [CII] emitter at $z=5.955$ or a low-$z$ CO emitting galaxy, this result encourages us to search for such mm-wave line emitters using the ALMA data. This may allow us to constrain [CII] and/or CO luminosity functions across the history of the universe (e.g., \cite[Ono et al.~2014]{Ono2014}, \cite[Tamura et al.~2014]{Tamura2014}, \cite[Matsuda et al.~2015]{Matsuda2015}).

\underline{\it Implications for future ALMA deep surveys:} We find a rapid increase in the number of faint ALMA sources below $\sim$0.3 mJy as expected by the latest source counts around 1 mm, demonstrating that undertaking a shallower survey by a factor of $\sim$2 would drastically reduce the detections (only 3 sources above 4$\sigma$). Our recent deep ALMA band-6 survey of the central 4.5 arcmin$^2$ region of the SSA22 field (which exhibits an extreme over-density at $z=3.1$, \cite[Tamura et al.~2009]{Tamura2009}) at a similar depth ($1\sigma$ 0.07 mJy at 1.1 mm; \cite[Umehata et al.~2015]{Umehata2015}) also supports this view. It is also noteworthy to mention the rapid progress of new large single dishes such as LMT 32 m (to be extended to 50 m) with AzTEC. Recently it demonstrates a survey depth down to 0.17 mJy (1$\sigma$) at 1.1 mm to constrain the rest-FIR properties of a $z_{\rm photo}=9.6$ galaxy (\cite[Zavala et al.~2015]{Zavala2015}), suggesting that much deeper depths will ensure a unique parameter space that exploits the unique capabilities of ALMA. 

\smallskip

This makes use of the following ALMA data: ADS/JAO.ALMA\#2012.1.00756.S.

\end{document}